# Genome-wide association and transcriptome analysis reveals serum ghrelin to be linked with GFRAL

*Short title: GWA and transcriptome analysis of serum ghrelin*

Dirk Alexander Wittekind,*[1] Markus Scholz,[3,4] Jürgen Kratzsch,[2] Markus Löffler,[3,4] Katrin Horn,[3,4] Holger Kirsten,[3,4] Veronica Witte,[5] Arno Villringer,[5] Michael Kluge*[1]

[1] Department of Psychiatry and Psychotherapy, University of Leipzig, Leipzig, Germany.

[2] Institute of Laboratory Medicine, Clinical Chemistry and Molecular Diagnostics, University of Leipzig, Leipzig, Germany.

[3] Institute for Medical Informatics, Statistics, and Epidemiology (IMISE), University of Leipzig, Leipzig, Germany.

[4] LIFE Research Center for Civilization Diseases, University of Leipzig, Leipzig, Germany

[5] Department of Neurology, Max Planck Institute for Cognitive and Brain Sciences, Leipzig, Germany.

***Corresponding authors:**

Department of Psychiatry and Psychotherapy
University of Leipzig
Semmelweisstrasse 10
D-04103 Leipzig
Germany

DAW: Email: dirkalexander.wittekind@medizin.uni-leipzig.de - Tel.: +49 341 97 20467
MK: Email: michael.kluge@medizin.uni-leipzig.de - Tel.: +49 341 97 24673

## Disclosure summary

The authors have nothing to disclose

**Abstract**

Objective: Ghrelin is an orexigenic peptide hormone involved in the regulation of energy homeostasis, food intake and glucose metabolism. Serum levels increase anticipating a meal and fall afterwards. Underlying genetic mechanisms of the ghrelin secretion are unknown.

Methods: Total serum ghrelin was measured in 1501 subjects selected from the population-based LIFE-ADULT-sample after an overnight fast. A genome-wide association study (GWAS) was performed. Gene-based expression association analyses (transcriptome-wide association study (TWAS)) were done using MetaXcan.

Results: In the GWAS, three loci reached genome-wide significance: the WW-domain containing the oxidoreductase-gene (WWOX; p=1.80E-10) on chromosome 16q23.3-24.1 (SNP: rs76823993); the Contactin-Associated Protein-Like 2 gene (CNTNAP2; p=9.0E-9) on chromosome 7q35-q36 (SNP: rs192092592) and the Ghrelin And Obestatin Prepropeptide gene (GHRL; p=2.72E-8) on chromosome 3p25.3 (SNP: rs143729751). In the TWAS, serum ghrelin was negatively associated with RNA expression of the GDNF Family Receptor Alpha Like (GFRAL), receptor of the anorexigenic Growth Differentiation Factor-15 (GDF15), (z-score=-4.288, p=1.81E-05). Furthermore, ghrelin was positively associated with Ribosomal Protein L36 (RPL36; z-score=4.848, p=1.25E-06).

Conclusions: Our findings provide evidence of a functional link between two major players of weight regulation, the ghrelin system and the GDF15/GFRAL-pathway.

## 1. **Introduction**

Ghrelin is a 28-amino-acid peptide hormone synthesized in the stomach.[1] Ghrelin binds to its receptor, the growth hormone secretagogue receptor 1a (GHSR1a), after being acylated by ghrelin-O-acyl-transferase (GOAT).[2,3] Ghrelin is secreted in a pulsatile manner and factors regulating ghrelin serum levels are circadian rhythm, sex, feeding status, body fat and stress.[4,5] However, underlying genetic mechanisms of the ghrelin secretion are unknown. Ghrelin receptors and GOAT are broadly expressed, including in intestine, pituitary, kidney, lung, heart, pancreatic islets, endocrine tissue and the central nervous system (CNS).[6] Accordingly, ghrelin has been shown to be involved in various biological functions. It stimulates the hypothalamic pituitary (HP)-adrenal and the growth hormone axes,[1] but inhibits the HP-gonadal and the HP-thyroid axes (TSH).[7,8] In the CNS, ghrelin was shown to activate the reward system,[9,10] increasing food, drug and alcohol consumption, being considered a relevant factor for the development of addiction.[10,11] Published evidence consistently points to anxiolytic and antidepressant effects and a prominent role in learning, memory and neuroprotection.[11,12] It increases blood glucose by suppressing pancreatic insulin secretion, increases hepatic gluconeogenesis and stimulates gastric motility and emptying as well as lipogenesis.[5,13] While ghrelin has been implicated in various cancers, its precise role in cancer development and progression is unclear.[14]

Ghrelin increases weight by reducing energy expenditure and increasing appetite and thereby food intake.[15] The orexigenic (appetite increasing) effect is mediated at hypothalamic level by stimulating neurons containing orexigenic neuropeptide Y (NPY) and agouti-related protein (AgRP) and by inhibiting neurons containing anorexigenic a-melanocyte-stimulating hormone (a-MSH) and cocaine- and amphetamine-regulated transcript (CART).[5,16,17] Leptin, an important adipocyte-derived *an*orexigenic hormone, exerts opposite effects on these neurons.[18] GDF15 is another anorexigenic peptide with a different mode of action.[19,20] GDF15 binds to GFRAL located in the area postrema.[21–24] The clinical relevance of the GDF15/GFRAL-pathway, e.g. in mediating metformin’s` metabolic effects, is just being recognized.[20] There has been almost no information on the association between GDF15/GFRAL-pathway and the ghrelin system so far.

Due to its involvement in various auto-regulatory systems, the ghrelin system has been identified as a potential drug target for several conditions including tumor cachexia[25] and addiction with promising results in recent years.[26]

In order to elucidate genetic mechanisms underlying the ghrelin secretion, a first GWAS of total ghrelin in serum and a TWAS were performed, integrating expression quantitative trait loci (eQTL) information across various tissues with GWAS results using MetaXcan.[27]

## 2. **Methods**

### 2.1 *Study design and subjects*

Subjects were recruited in the framework of the LIFE-Adult study, a population-based cohort with 10,000 adults (age range mainly 40-79 years, 400 subjects with an age range of 18-39 years). Participants were age- and sex-stratified randomly recruited in the city of Leipzig, Germany (for details of the study design see).[28] Total ghrelin was measured in 1666 subjects. From these, 165 subjects were excluded due to an incomplete data set, i.e. when any of the parameters required for data analyses was missing such as body mass index (BMI) or genetic data of sufficiently high quality. Thus, the resulting study population comprised 1501 subjects (807 men and 694 women). All participants gave written informed consent to take part in the study. The procedures were conducted according to the Declaration of Helsinki and approved by the ethics committee of the University of Leipzig (registration-number: 263-2009-14122009).

*2.2 Ghrelin measurements*

Blood samples were collected after an overnight fast between 07:30 and 10:30 h, serum was separated by centrifugation and then frozen and stored at -80°C. Ghrelin in serum was measured using a radioimmunoassay for total ghrelin (Mediagnost, Reutlingen Germany). Sensitivity of the assay was 0.04 ng/mL, mean intra-assay coefficients of variation were 2.7-4.3%; inter-assay coefficients of variation were between 6.9% and 9.2% for the mean expected range of clinical data around 0.88 and 0.97 ng/mL.

*2.3 Genotyping, preprocessing and imputation*

Subjects were genotyped with the genome-wide SNP array Axiom Genome-Wide CEU 1 Array Plate (Affymetrix, Santa Clara, California, USA). Details of genotyping and primary quality control of samples and SNPs can be found elsewhere.[29] In brief, sample quality control included analysis of dishQC, call rate, heterozygosity, sex mismatches, cryptic relatedness and for X-chromosomal analysis irregularities of X-Y intensity plots. SNP quality control included call rate (≤97%), parameters of cluster plot irregularities as suggested by Affymetrix best practice, violation of Hardy-Weinberg equilibrium (p≥$10^{-6}$ in exact test) and plate associations (p≥$10^{-7}$). In total 4985 (4978) samples and 532875 (13554) SNPs fulfilled all quality criteria for autosomal (X-chromosomal respectively) analysis.

Genotypes of the 1000 Genomes reference phase 3 , version 5[30] were imputed using IMPUTE2 (version v2.3.2)[31] after pre-phasing with SHAPEIT (version v2.r837).[32] SNPs with low imputation quality (IMPUTE info score<0.5) and bad power (minor allele frequency<2%) were filtered. A total of 9,868,623 SNPs were considered for analysis.

*2.4 GWAS analyses*

Ghrelin, covariables and genotype data are available for a total of 1501 samples. Association analysis was performed with SNPTEST (version 2.5.2 ). Logarithmized ghrelin levels were adjusted for age, sex, alcohol intake, smoking status and logarithmized BMI and an additive model of inheritance was assumed. X-chromosomal variants were analyzed assuming total X-inactivation. A-value cut-off of $5x10^{-8}$ was considered genome-wide significant. Independent SNPs were determined by priority pruning applying a linkage disequilibrium cut-off of $r^2 \geq 0.5$.

SNPs were comprehensively annotated by an in house pipeline as explained elsewhere.[33] In brief, annotation comprised nearby genes using Ensemble, LD based look-up of other GWAS traits presented in the GWAS Catalog,[34] and of expression quantitative trait loci (eQTLs) as reported by GTex and own blood data.[35]

*2.5 TWAS Analysis*

For gene-based expression association analyses, we utilized MetaXcan.[27] We applied the GTEXv7 models, MESA,[36] and DGN-model as it was trained by the authors. For this analysis, we used the same SNP filter criteria regarding MAF and info-score as in our main GWAS. Correction for multiple testing for the number of tested tissue and tested genes was done at the FDR 50% level in a hierarchical way considering genes as the unit of interest as previously described.[37]

## 3. Results

*3.1 GWAs analysis*

GWAS was performed in the population-based ‘LIFE-Adult study’. Sample description is provided in Table 1. GWA analysis revealed no signs of general inflation of test statistics (Lambda=1.00). Independent SNPs reached genome-wide significance and could be attributed to three distinct genetic loci.

The strongest hit was an intronic variant of the WW-domain containing the oxidoreductase-gene (WWOX) located on chromosome 16q23.3-24.1 (SNP: rs76823993; MAF=1.7%, explained variance 2.4%, p=1.8E-10.) Carriers of the minor allele had lower ghrelin serum levels (Fig. 1, Table 2). Six further SNPs in the WWOX gene support the signal (p<E-6) (Fig. 2).

The second strongest genetic variant reaching genome-wide significance is located on chromosome 7q35-q36.1 in the Contactin-Associated Protein-Like 2 gene (CNTNAP2; SNP: rs192092592; MAF=1.2%, explained variance 2.0%, p=9.0E-9) (Fig. 1, Table 2). However, this SNP is not supported by other variants, requiring future validation. Carriers of the minor allele showed lower ghrelin serum levels.

The third genetic variant reaching genome-wide significance was on the Ghrelin And Obestatin Prepropeptide (GHRL) gene in the position 3p25.3 itself which is biologically highly plausible (SNP: rs143729751; MAF=2.1%, explained variance 1.8%, p=2.72E-8). The SNP is in some linkage disequilibrium (LD) with an eQTL of GHRL in blood (r2=0.64). Carriers of the minor allele showed lower ghrelin serum levels (Fig. 1, Table 2). 14 hits in 9 different genes reached trend significance (Table 2).

*3.2 TWAs analysis*

In a MetaXcan-analysis, ghrelin serum levels were associated with messenger RNA (mRNA) expression of the GFRAL gene in adipose tissue (z-score=-4.288; p=1.81E-05, False-discovery-rate (FDR)=0.147, number of SNPs used=109). In addition, ghrelin serum levels were also associated with messenger RNA (mRNA) expression of the RPL36 gene (z-

score=4.848; p=1.25E-06, FDR=0.011, number of SNPs used=20) in a variety of tissues including cerebellum, whole blood, subcutaneous fat, esophagus mucosa, skeletal muscles and transverse colon. That means that a lower expression of GFRAL and a higher expression of RPL36 were associated with higher levels of serum ghrelin. Regional association plots summarize the evidence of all eQTLs of GFRAL and RPL36 that were used to identify the gene-expression association with serum ghrelin (Fig. 3).

## 4. Discussion

In this GWAS total ghrelin serum levels as trait were investigated. We were able to identify three genetic loci reaching genome-wide significance and 14 in 9 genes that showed an association with ghrelin serum levels at a trend level. The strongest and best supported hit was an intronic variant in the WWOX-gene (p=1.80E-10). WWOX spans the second most common fragile site FRA16D and encodes for a 414-amino acid protein that contains two WW domains. WW structures are known to be involved in protein-protein interactions. *WWOX* is highly expressed in hormonally regulated tissues (testis, prostate, and ovary).[38]

The second strongest hit was an intronic variant in the CNTNAP2-gene (p=9.0E-9). It is the longest gene in the human genome, encompassing almost 1.5% of chromosome 7. It encodes for the protein Contactin-Associated Protein-Like 2 (CASPR2). CASPR2 is part of the neurexin superfamily and functions as cell adhesion molecules and receptors in the central nervous system of vertebrae. It is mainly localized at the juxtaparanodes of myelinated axons and mediates interactions between neurons and glia during nervous system development. It is also involved in localization of potassium channels within differentiating axons.[39]

The third hit was in the GHRL-gene itself (p=2.72E-8). GHRL encodes for a 117-amino acid preprohormone called preproghrelin. From this, the 28-amino acid des-acyl ghrelin is spliced, as well as obestatin and the C-terminal ending of the preproghrelin, called C-ghrelin. Obestatin was originally believed to be a ghrelin-antagonist, but more recent evidence has questioned this and both its exact function and the function of C-ghrelin remain unclear.[5]

All candidate genes are functionally highly plausible:

When looking at the biological functions coded by WWOX and CNTNAP2, there is a striking overlap with functions of ghrelin, mainly concerning processes in the CNS. Both WWOX and CNTNAP2 seem to be involved in neuronal development, branching and maturation.[40,41] CNTNAP2 is involved in language processing and development in both animals and humans and a multitude of studies shows mental retardation and speech impairment in children with genetic aberrations in both genes as well as strong associations with epilepsy.[42] This is of relevance, as ghrelin has been consistently shown to exert neuroprotective and anticonvulsant effects and promote neurogenesis, mainly in the hippocampus.[11] WWOX and CNTNAP2 are associated with Alzheimer's disease, with both being identified as risk loci for late onset Alzheimer disease (LOAD) in Alzheimer's disease GWAs and GWA-meta-analyses.[43,44]

Genetic variants of WWOX have been linked to obesity, type 2 diabetes, high fasting glucose and high body-mass-index,[45] thus showing a strong overlap with ghrelin function. A loss of function of WWOX has been documented in many tumor entities[46] and WWOX has been shown to be a tumor-suppressor gene.[47] Ghrelin is expressed in many cancers like renal cell carcinoma, pancreatic cancer, thyroid cancer, lung cancer, breast cancer, prostate cancer, gastric cancer and colorectal carcinoma. Its role in cancer growth and progression is being controversially discussed, e.g. due to a presumably dosage-dependent effect on tumor growth.[14]

CNTNAP2 has been repeatedly linked to psychiatric diseases. Especially for autism, there is broad evidence suggesting an involvement of CNTNAP2 in the pathogenesis of this disease.[48,49] CNTNAP2 has also been shown to be associated with schizophrenia, bipolar disorder and major depression.[39,50] Ghrelin has been consistently shown to exert antidepressant and anxiolytic properties in animals models and the ghrelin system has been repeatedly shown to be altered in patients suffering from major depression.[11] Also the fact that auto-antibodies against CASPR2, CNTNAP2's gene product, can cause autoimmune-encephalitis often presenting with schizophrenia-like symptomatology is in line with CNTNAP2's involvement in psychiatric illnesses.[51]

In a TWAS using MetaXcan,[27] exploiting tissue-specific expression quantitative trait loci (eQTLs), a negative association between the expression of the GFRAL gene in subcutaneous adipose tissue and ghrelin serum levels was found. The GDF15/GFRAL-system has been identified recently. In 2017, four groups reported that brain-stem located GFRAL, detected in 2005 as an orphan receptor,[52] is the receptor for GDF15, requiring receptor tyrosine kinase (RET) as a co-receptor.[21–24] GDF15, also known as Macrophage Inhibitory Cytokine-1 (MIC-1), had been identified in 2007 as a peptide mediating anorectic/cachectic effects.[19] Its relevance is just being recognized: For example, weight loss associated with metformin, the worldwide most prescribed antidiabetic, was shown to correlate with increase of GDF15 and to depend on the integrity of the GDF15/GFRAL system.[20]

Our findings suggest that increased expression of GFRAL in subcutaneous adipose tissue has a directional effect on reducing ghrelin serum levels. This is very plausible, as the GDF15/GFRAL-system and the ghrelin system have opposing effects in the organism, not only in energy metabolism and appetite control. A further parallel between ghrelin and GDF15, is that both are produced in times of stress.[5,53,54] Here, too, they have opposite effects on appetite and food intake, being in line with a metabolic interaction also in stress. In addition, very recently, GDF-15 was shown to have pro-emetic effects,[55] while ghrelin is known to be a strong anti-emetic agent and promoter of gastric motility.[56] This further points to an antagonistic relationship between ghrelin and GDF-15/GFRAL. Finally, for ghrelin and GDF15 both tumor-growth promoting and -inhibiting effects are being discussed.[14,57]

Yet, while the presence of GFRAL-mRNA in subcutaneous adipose tissue is a robust and replicated finding,[23,27] its biological relevance in subcutaneous tissue remains to be elucidated, since its protein product, the GFRAL-receptor, has been detected so far only in the brain stem.[22,23]

So far, no data exists linking GFRAL and ghrelin and only very scarce data studying GDF15 and ghrelin. Here, a non-peptidergic ghrelin receptor agonist did not affect GDF15 in mice.[58]

A positive association with ghrelin serum levels was found in the TWAS also for mRNA expression levels of the RPL36 gene in various tissues including cerebellum, whole blood, subcutaneous fat, esophagus mucosa, skeletal muscles and transverse colon. RPL36 is a subunit of the 60S ribosomal protein and as such involved in protein synthesis and cell proliferation.[59] This finding is biologically plausible as ghrelin functions as an anabolic hormone. Furthermore, RPL36 has been implicated like ghrelin in carcinogenesis with a so far unclear role.[59]

Limitations of this study are the relatively small case sample for a GWAS and that no replication cohort was available.

In conclusion, we identified three functionally plausible genetic loci reaching genome-wide significance in the GWAS of ghrelin serum levels. Furthermore, performing a TWAS, we could link the ghrelin system with the GDF15/GFRAL-system, suggesting an interaction between these systems having opposite effects on appetite control and energy homeostasis. Further research based on these findings is warranted as they might lead to a better understanding of and innovative treatment approaches for metabolic, oncological and neuropsychiatric diseases.

**Funding**

This work was supported by LIFE − Leipzig Research Center for Civilization Diseases, University of Leipzig. LIFE is funded by means of the European Union, by means of the European Social Fund (ESF), by the European Regional Development Fund (ERDF), and by means of the Free State of Saxony within the framework of the excellence initiative.

**Acknowledgements**

None

## Figures

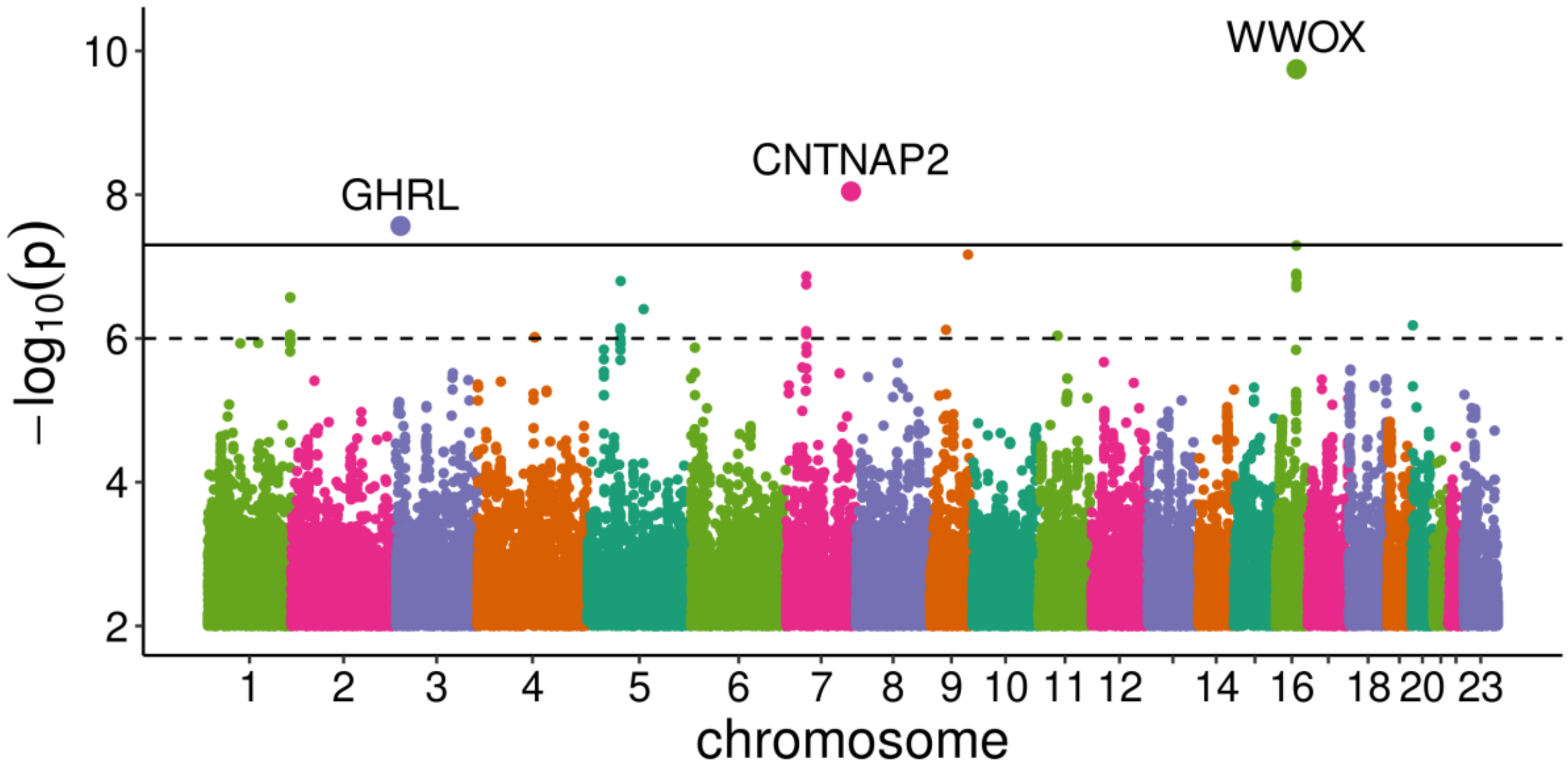

**Fig. 1: Manhattan plot showing the SNP associations for total ghrelin serum levels**. The limit for genome-wide significance ($p=5.0x10^{-8}$) is displayed as line.

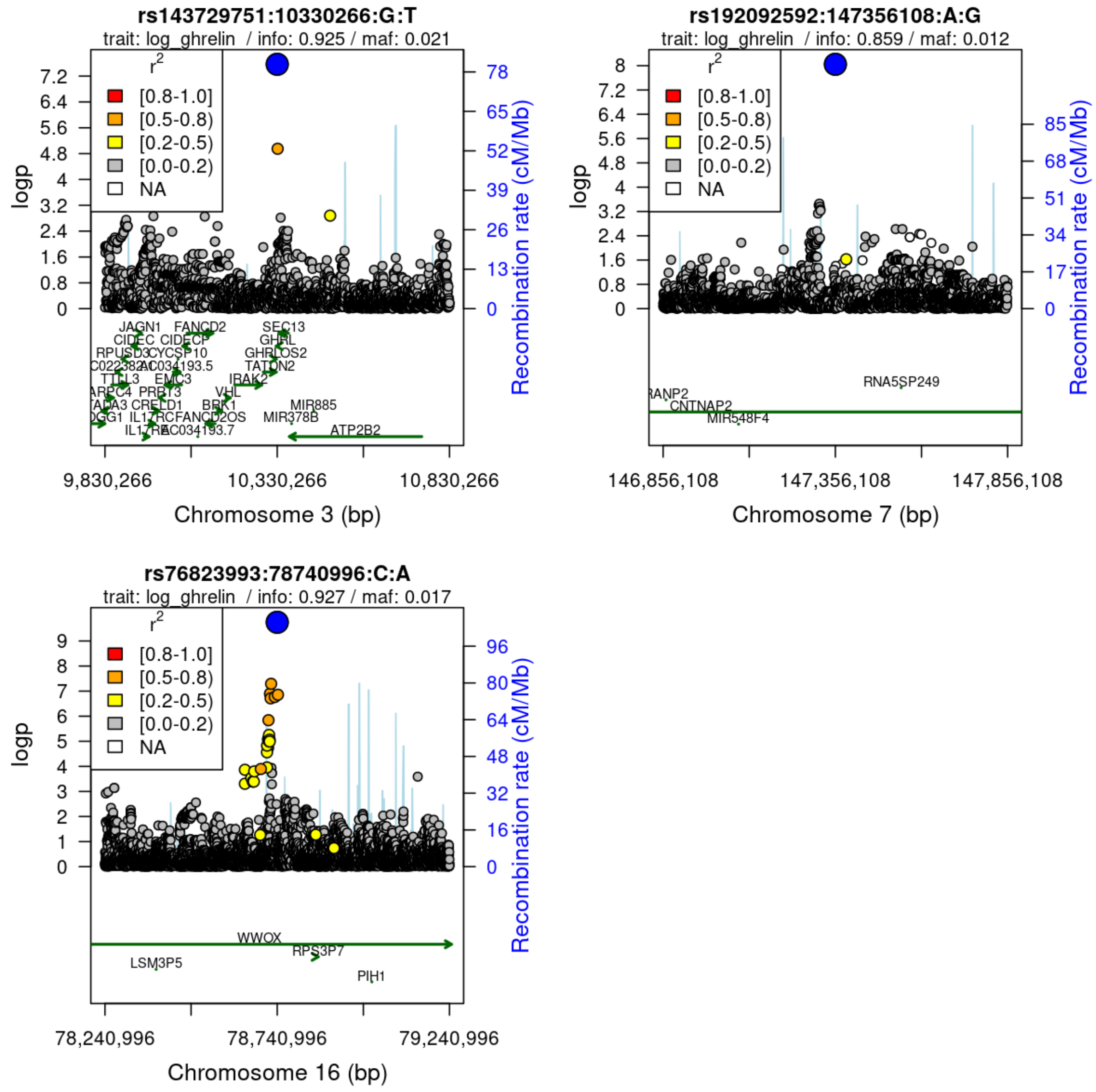

**Fig. 2: Regional association (RA) plots of three GWAS hits.** The lead SNP is colored blue, and the other SNPs are colored according to their LD with the lead SNP (using 1000 Genomes Phase 3, Europeans only). (A) RA plot for rs143729751 and GHRL at cytoband 3p25.3. (B) RA plot for rs192092592 and CNTNAP2 at 7q35. (C) RA plot for rs76823993 and WWOX at 16q23.1.

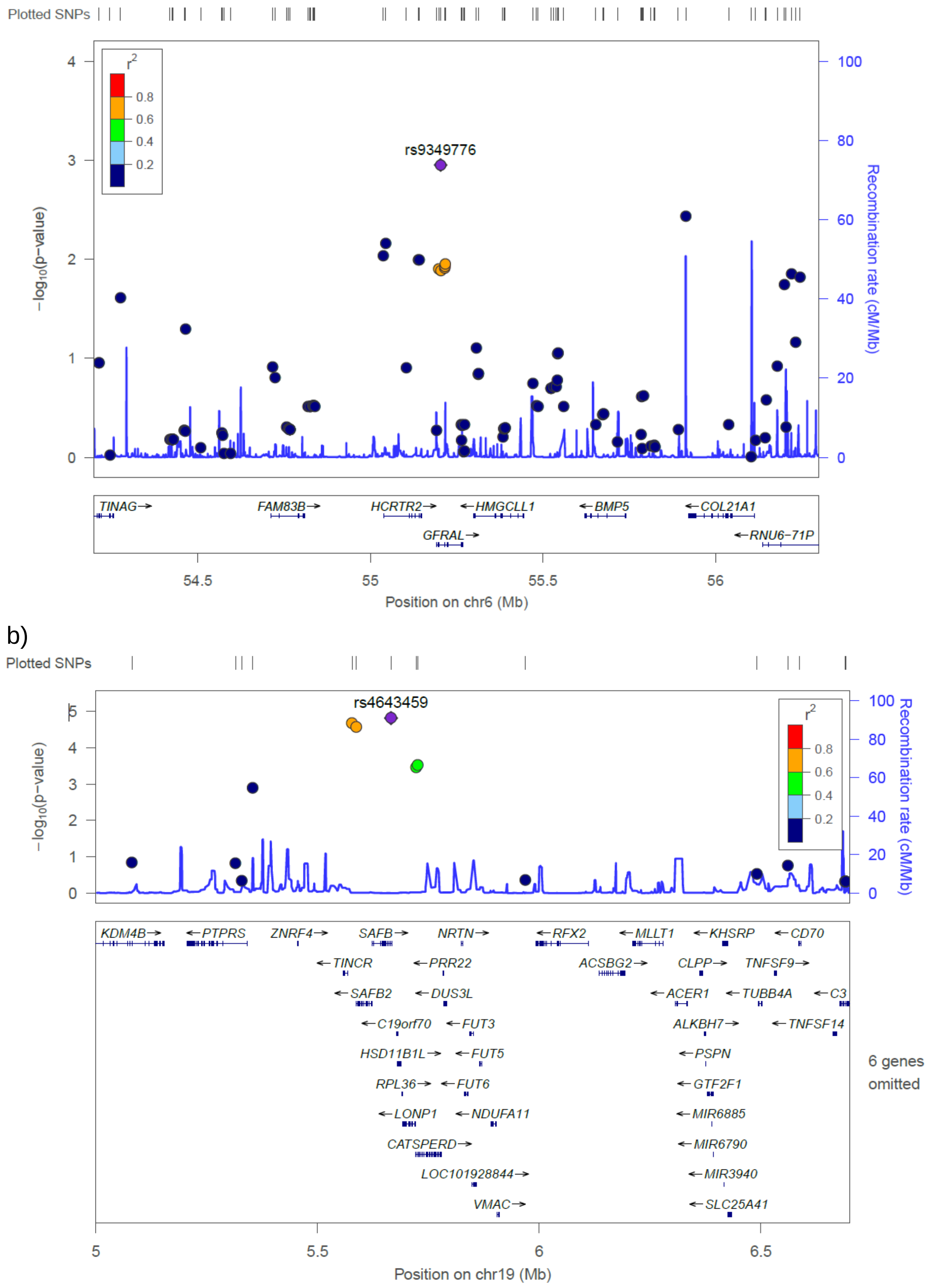

**Fig. 3: Regional association plots of GFRAL (a) and RPL36 (b)** showing all eQTL-SNPs included in the gene expression model. It illustrates that many only nominal association of SNPs with ghrelin contribute jointly to the significant gene-level association of GFRAL (a) respectively RPL36 (b) with ghrelin.

## Tables

**Table 1:** Description of the study sample

| | **Total sample (n=1501)** | **Males (n=807) (53.8%)** | **Females (n=694) (46.2%)** | ***p-value*** |
|---|---|---|---|---|
| **Age,** years, mean (SD) | 57.9 (15.2) | 57.5 (15.5) | 58.3 (14.7) | 0.78[a] |
| **BMI,** kg/m$^2$, mean (SD) | 27.1 (4.5) | 27.4 (4.0) | 26.7 (4.9) | **<0.001*****[a] |
| **Non-smokers,** n (%) (n=1457) | 1235 (84.8%) | 651 (82.4%) (n=790) | 584 (87.6%) (n=667) | **0.006****[b] |
| **Ghrelin serum,** pg/ml, mean (SD) | 927.6 (450.6) | 812.9 (306.7) | 1061.0 (545.0) | **<0.001*****[a] |

[a] A Kolmogorov-Smirnov-Test was performed in order to assess normal distribution. Because of non-normal distribution of the dependent variables according to these results, a Mann-Whitney-U-test was applied to compare ghrelin serum levels (Z=-11.13), age (Z=-0.29) and BMI values (Z=-4.19) between sexes. [b] A chi$^2$ test for two-by-two tables was used.

**Table 2: Loci showing genome-wide significance and trend significance for total ghrelin serum levels**

**Loci showing genome-wide significance (*P* <5E-6)**

| SNP | Cytoband | Chr | Position | Candidate genes | Impute Info score | Effect allele | Other allele | MAF (Effect allele) | Beta (CI) (Effect allele) | *P* | $\eta^2$ |
|---|---|---|---|---|---|---|---|---|---|---|---|
| rs76823993 | 16q23.1 | 16 | 78740996 | WWOX | 0.927 | A | C | 0.017 | -0.382 (-0.5 - -0.27) | 1.80E-10 | 0.024 |
| rs192092592 | 7q35 | 7 | 147356108 | CNTNAP2 | 0.859 | G | A | 0.012 | -0.439 (-0.59 - -0.29) | 9.00E-9 | 0.019 |
| rs143729751 | 3p25.3 | 3 | 10330266 | GHRL | 0.952 | T | G | 0.021 | -0.299 (-0.4 - -0.19) | 2.72E-8 | 0.018 |

**Loci showing trend significance (*P* <1E-6)**

| | | | | | | | | | | | |
|---|---|---|---|---|---|---|---|---|---|---|---|
| rs139359241 | 9q33.1 | 9 | 121992761 | BRINP1 | 0.834 | A | G | 0.015 | -0.36 (-0.5- -0.24) | 6.82E-8 | 0.017 |
| rs117196347 | 16q23.1 | 16 | 78720394 | WWOX | 0.988 | T | C | 0.026 | -0.25 (-0.34- -0.16) | 1.26E-7 | 0.017 |
| rs118158533 | 16q23.1 | 16 | 78720391 | WWOX | 0.949 | C | G | 0.016 | -0.32 (-0.44- -0.20) | 1.30E-7 | 0.017 |
| rs74483218 | 7p14.1 | 7 | 38031039 | SFRP4 | 0.902 | A | G | 0.015 | -0.35 (-0.48- -0.22) | 1.37E-7 | 0.016 |
| rs199653320 | 5q12.3 | 5 | 65124780 | NLN | 0.997 | AC | A | 0.010 | -0.40 (-0.55- -0.25) | 1.59E-7 | 0.016 |
| rs139200008 | 16q23.1 | 16 | 78734010 | WWOX | 0.917 | CTA | C | 0.026 | -0.25 (-0.35- -0.16) | 1.71E-7 | 0.016 |
| rs112426408 | 1q44 | 1 | 244052447 | AKT3 | 0.87 | G | A | 0.029 | -0.23 (-0.33- -0.15) | 2.68E-7 | 0.016 |
| rs77563704 | 5q14.3 | 5 | 91845530 | RP11-133 | 0.936 | A | C | 0.050 | -0.17 (-0.24- -0.11) | 3.92E-7 | 0.015 |
| rs187860960 | 20p12.3 | 20 | 5920093 | TRMT6 | 0.84 | T | C | 0.025 | -0.25 (-0.35 - -0.15) | 6.58E-7 | 0.015 |
| rs138296128 | 5q12.3 | 9 | 36885124 | PAX5 | 0.827 | A | C | 0.012 | -0.38 (-0.53- -0.23) | 7.60E-7 | 0.015 |
| rs113194124 | 7p14.1 | 7 | 38014091 | SFRP4 | 0.965 | T | C | 0.027 | -0.23 (-0.33- -0.14) | 7.94E-7 | 0.014 |
| rs80240706 | 1q44 | 1 | 243997795 | AKT3 | 0.901 | T | C | 0.027 | -0.23 (-0.32- -0.14) | 8.87E-7 | 0.014 |
| rs142653572 | 11p11.2 | 11 | 44370951 | ALX4 | 0.818 | T | C | 0.011 | -0.39 (-0.54- -0.23) | 9.21E-7 | 0.014 |
| rs142224718 | 4q24 | 4 | 102643916 | BANK1 | 0.900 | C | T | 0.011 | -0.37 (0.53- -0.23) | 9.64E-7 | 0.014 |

SNP: Single nucleotide polymorphism, Chr: Chromosome, Info: information quality of imputed SNPs according to IMPUTE 2, Min/Maj: Minor allele/major allele, MAF: Minor allele frequency, Beta (CI): Beta coefficient (confidence interval), $\eta^2$= explained variance, WWOX: WW-domain containing oxiduoreductase-gene CNTNAP 2: Contactin Associated Protein Like 2, GHRL: Ghrelin And Obestatin Prepropeptide, BRINP1: BMP/retinoic acid inducible neural specific 1, SFRP4: Secreted frizzled related protein 4, NLN: Neurolysin, AKT3: AKT Serine/Threonine Kinase 3, RP11-133: long non-coding RNA, manual annotation from Havana project, TRMT6: TRNA Methyltransferase 6, PAX 5: Paired Box 5, ALX4: ALX Homeobox 4, BANK1: B Cell Scaffold Protein With Ankyrin Repeat